\documentstyle[12pt]{article}

\newcommand{\ba}{\begin{array}}
\newcommand{\ea}{\end{array}}

\newcommand{\be}{\begin{equation}}
\newcommand{\ee}{\end{equation}}
\newcommand{\bea}{\begin{eqnarray}}
\newcommand{\eea}{\end{eqnarray}}
\newcommand{\beal}{\setcounter{letter}{1} \begin{eqnarray}}
\newcommand{\eeal}{\addtocounter{equation}{1} \end{eqnarray}}
\newcommand{\none}{\nonumber \\}
\newcommand{\req}[1]{Eq.(\ref{#1})}

\newcommand{\larrow}{\,\,\,\,\hbox to 30pt{\rightarrowfill}
\,\,\,\,}
\newcommand{\slarrow}{\,\,\,\hbox to 20pt{\rightarrowfill}
\,\,\,}
\newcommand{\oli}[1]{\overline{#1}}

\newcommand{\der}{\partial}

\newcommand{\half}{{1\over2}}

\newcommand{\gcal}{{\cal G}}
\newcommand{\fcal}{{\cal F}}
\newcommand{\jcal}{{\cal J}}

\newcommand{\gtilde}{\tilde{\cal{G}}}

\newcommand{\jos}{\frac{\der^2 \phi}{\der {z_{+}}^2}}
\newcommand{\jon}{\frac{\der^2 \phi}{\der {z_{-}}^2}}
\newcommand{\joc}{\frac{\der^2 \phi}{\der z_{+} \der z_{-}}}
\newcommand{\mac}{\frac{\der^2 \rho}{\der z_{+} \der z_{-}}}
\newcommand{\drozp}{\frac{\der \rho}{\der z_{+}}}
\newcommand{\drozn}{\frac{\der \rho}{\der z_{-}}}
\newcommand{\dfizp}{\frac{\partial \phi}{\partial z_{+}}}
\newcommand{\dfizn}{\frac{\der \phi}{\der z_{-}}}

\newcommand{\Vbar}{\oli{V}}

\newcommand{\gbar}{\oli{g}}
\newcommand{\phibar}{\oli{\phi}}
\newcommand{\Abar}{\oli{A}}
\newcommand{\Wbar}{\oli{W}}
\newcommand{\Fbar}{\oli{F}}
\newcommand{\deltabar}{\oli{\delta}}

\newcommand{\prho}{\hat{\Pi_\rho}}
\newcommand{\pphi}{\hat{\Pi_\phi}}
\newcommand{\pa}{\hat{\Pi}_{A_1}}
\begin{document}
\hsize =6.0 in
\vsize =11.7 in
\hoffset=-0.1 in
\voffset=-0.5 in
\baselineskip=15pt
\mbox{}\\[1cm]
\begin{center}
{\bf TWO DIMENSIONAL DILATON GRAVITY COUPLED TO\\
 AN ABELIAN GAUGE FIELD}\\
\vspace{0.5 cm}
{\sl by\\
}
\vspace*{0.40cm}
{\bf
D. Louis-Martinez ${}^{\sharp\,\flat}$ and G. Kunstatter ${}^\sharp$
\\}
\vspace*{0.40cm}
{\sl
$\sharp$ Dept. of Physics, University of Winnipeg and\\
Winnipeg Institute of
Theoretical Physics\\
Winnipeg, Manitoba, Canada R3B 2E9\\
{[e-mail: gabor@theory.uwinnipeg.ca]}\\[5pt]
}
{\sl
$\flat$ Dept. of Physics, University of Manitoba\\
Winnipeg, Manitoba, Canada R3T 2N2\\
{[e-mail: martinez@uwpg02.uwinnipeg.ca ]}
}
\end{center}
\bigskip\noindent
{\large
ABSTRACT
}
\par
\noindent
The most general two-dimensional dilaton gravity theory  coupled to an Abelian
gauge field is
considered. It is shown that, up to spacetime diffeomorphisms and $U(1)$ gauge
transformations,
the field equations admit a two-parameter family of distinct, static solutions.
 For theories with
black hole solutions, coordinate invariant expressions are found for the
energy, charge, surface
gravity, Hawking temperature and entropy of the black holes. The Hawking
temperature is
proportional to the surface gravity as expected, and both vanish in the case of
extremal black holes
in the generic theory. A Hamiltonian analysis of the general theory is
performed, and a complete
set of (global) Dirac physical observables is obtained. The theory is then
quantized using the Dirac
method in the WKB approximation. A connection between the black hole entropy
and the imaginary part of the WKB phase of the Dirac quantum wave functional is
found for arbitrary
values of the mass and $U(1)$ charge. The imaginary part of
the  phase vanishes for extremal black holes and for eternal, non-extremal
Reissner-Nordstrom black holes.\\[10pt]
arch-ive/9503016
\newpage
\renewcommand{\baselinestretch}{1.5}
\section{Introduction}\medskip
\par
Two dimensional models of gravity have  been the subject of much research
in the hope that the simplicity of these models might yield clues to
theoretical problems associated with the
evaporation of black holes via Hawking radiation\cite{info}. One question of
particular current
interest concerns the nature of the radiation for charged black holes near
extremality where the
standard thermal description is expected to break down\cite{extreme}. Recently,
a unified
description was presented of the most general 2-D  vacuum dilaton gravity
theory\cite{gkm}: the
Killing vector for the generic theory was found, and in the case where black
hole solutions were
present, expressions for the surface gravity and entropy were derived. In the
process, an intriguing
relationship was uncovered between the black hole entropy and the phase of the
Dirac quantized
physical wave functionals for stationary states in the theory.
\par
The purpose of the present paper is to extend the analysis of \cite{gkm} to
include coupling to a
$U(1)$ gauge field. The action functional we will consider depends on the
two-dimensional metric
tensor, the dilaton scalar and an Abelian gauge field. We consider the most
general case for which
the field equations are at most second order in derivatives. This allows two
arbitrary functions of
the dilaton field  in the action.\footnote{The Dirac quantization of dilaton
gravity coupled to a
Yang-Mills field has recently been considered by Strobl\cite{strobl}, while the
perturbative
quantization of Dilaton-Maxwell theory has been examined by Elizalde and
Odinstov\cite{odintsov}.} The first is the dilaton potential, while the second
is effectively a
dilaton dependent ``electromagnetic coupling" in the Maxwell term for the gauge
field.   For
specific choices of these two functions, we obtain various models of current
interest. One such
special case describes  ``dimensionally reduced"  3+1 spherically symmetric
black holes\cite{ssg} with electric charge.
Another interesting example is Jackiw-Teitelboim 2-D
gravity\cite{JT} with $U(1)$ gauge coupling, which provides a dimensionally
reduced model\cite{ortiz} for the spinning (axially symmetric) black holes of
Ba\~nados, Teitelboim and
Zanelli\cite{BTZ}. String inspired dilaton gravity coupled to an
electromagnetic field, which was
discussed by Frolov\cite{frolov}, is also contained as a special case.
\par
The paper is organized as follows:
Section 2 presents the action and field equations.  Section 3 derives the most
general solution in
conformal gauge, identifying the two coordinate invariant parameters labelling
inequivalent
solutions. The Killing vector associated with each solution is also written
down in covariant form,
and used to derive a necessary condition for the existence of black hole
solutions. This condition
in effect  provides the equation of state for the black hole, from which the
thermodynamic
quantities can easily be derived. The Hamiltonian analysis is described in
Section 4. The reduced
phase space is shown to be four dimensional and explicit expressions are given
for the
corresponding (global) physical observables. Section 5 derives the
thermodynamical properties of
black hole solutions in the generic theory, including the  surface gravity and
entropy. It is shown
that the surface gravity vanishes for extremal black holes  in the generic
theory. The Dirac
quantization of the generic theory is presented in Section 6, and a
relationship is found between the
imaginary contribution to the WKB phase of stationary states and the entropy of
corresponding
classical black hole solutions.  The imaginary contribution to the phase
vanishes for extremal
black holes and for eternal, non-extremal Reissner-Nordstrom type black holes
in the general theory. In Section 7, specific cases of physical
interest are described
within this formalism. Finally, Section 8 closes with conclusions and prospects
for future work.
\par
\section{The Model}\medskip
The most general action functional depending on the metric tensor
$g_{\mu\nu}$, scalar field $\phi$ and vector potential $A_\mu$ in two spacetime
dimensions that  contains at most second derivatives of the fields can
be written\cite{odintsov}\cite{banks}:
\bea
S[g,\phi, A]&=&\int d^2x \sqrt{-g} \left[{1\over2G}\left( \half
g^{\alpha\beta}
\partial_\alpha \phi \partial_\beta \phi +{1\over
l^2} {V}(\phi) +
D(\phi)
R(g)\right) \right.\none
& &\left.-{1\over4} W(\phi) F^{\mu\nu}F_{\mu\nu}\right].
\label{eq: action 1}
\eea
where $R(g)$ is the Ricci curvature scalar and $F_{\mu\nu} =\partial_\mu A_\nu
- \partial_\nu
A_\mu$ is the Abelian field strength tensor. $V(\phi)$ and $W(\phi)$ are
arbitrary functions of the
dilaton  field. It should be noted that in the above, the fields $\phi$,
$g_{\mu\nu}$ and
$A_\mu$ are dimensionless, as is the 2-D Newton constant, $G$. This requires
the inclusion of a coupling constant, $l$, of dimension length in the
potential term.
\par
If $D(\phi)$ is a differentiable function of $\phi$ such that $D(\phi)\neq0$
and ${d D(\phi)\over
d \phi}\neq0$ for any admissable value of $\phi$ then  the
kinetic term for the scalar field can be eliminated by means of the
(invertible) field
redefinition\cite{domingo1}:
\bea
\gbar_{\mu\nu} &= &\Omega^2(\phi) g_{\mu\nu} \\
\phibar &=& D(\phi)\\
\Abar_\mu &=& A_\mu
\label{eq: field redefinitions}
\eea
where
\be
\Omega^2(\phi) = \exp \left( \half\int {d\phi \over (dD/d\phi)}\right)
\ee
In terms of the new fields, the action \req{eq: action 1} takes the form:
\be
S =  \int d^2x  \sqrt{-\gbar}\left[{1\over 2G} \left(\phibar
R(\gbar)+{1\over
l^2}\Vbar(\phibar)\right) -{1\over4} \Wbar(\phibar)
\Fbar^{\mu\nu}\Fbar_{\mu\nu}\right].
\label{eq: action 2}
\ee
where $\Vbar$ and $\Wbar$ are defined as:
\bea
\Vbar(\phibar) &=& {V(\phi(\phibar))\over \Omega^2(\phi(\phibar))}\\
\Wbar(\phibar) &=& {W(\phi(\phibar)) \Omega^2(\phi(\phibar))}
\eea
 For example in the Achuracco-Ortiz model\cite{ortiz}, $\Vbar= \Lambda \phibar$
$
(\Lambda>0)$ and $\Wbar=\phibar^3$, while in the string inspired
model\cite{frolov}, $\Vbar =
constant$ and $\Wbar = \phibar^2$. For spherically symmetric gravity(SSG) with
an electromagnetic field,
$\Vbar  =
1/\sqrt{2\phibar}$ and
$\Wbar = (2\phibar)^{3/2}$. In the following we consider the action in the form
\req{eq: action 2}, and henceforth drop the bars over the fields.
\par
The field equations that follow from the action \req{eq: action 2} are:
\be
R+{1\over l^2}{dV\over d\phi} - { G\over2}{dW\over d\phi}
F^{\alpha\beta}F_{\alpha\beta}
=0
\label{eq: field1}
\ee
\be
\nabla_\mu\nabla_\nu \phi- {1\over 2l^2} g_{\mu\nu} V(\phi) + G\left(
  \delta^\alpha_\mu\delta^\beta_\nu-{3\over4} g_{\mu\nu} g^{\alpha\beta}
\right) W(\phi) F_\alpha{}^\gamma F_\beta{}_\gamma=0
\label{eq: field2}
\ee
\be
\nabla_\nu \left(W(\phi) F^{\mu\nu}\right) =0
\label{eq: field3}
\ee
\par\medskip
\section{Generalized Birkhoff Theorem}
\par
We now provide a simple proof of the two-dimensional version of Birkhoff's
theorem for the
model, extending the results of \cite{domingo2}. Without loss of generality we
choose a
coordinate frame in which the metric is conformally flat, and introduce light
cone coordinates
$z_+=x+t$ and $z_-=x-t$. In this frame we have:
\bea
g_{++} &=& g_{--} = 0\\
g{+-}&=& g_{-+} =\half e^{2\rho(z_+, z_-)}\\
\eea
We also define for convenience $F(z_+,z_-):= 2 F_{+-} = - 2 F_{-+}$.  In
conformal gauge, the
field equations take the form:
\be
8 e^{-2\rho} \mac -{1\over l^2} { d V \over d \phi} -G e^{-4\rho} {d
W(\phi)\over d\phi} F^2 = 0
\label{a12}
\ee

\bea
 \jos +\jon -2\joc &-&2\drozp \dfizp -2\drozn \dfizn\nonumber\\
&+&\half
e^{2\rho} {V(\phi)\over l^2} + {G\over 2} e^{-2\rho} W(\phi)F^2= 0
\eea

\be
\jos + \jon + 2\joc - 2\drozp \dfizp - 2\drozn \dfizn -\half
e^{2\rho} {V(\phi)\over l^2}  - {G\over 2} e^{-2\rho} W(\phi)F^2= 0
\label{a14}
\ee

\be
\jos - \jon - 2\drozp \dfizp + 2\drozn \dfizn = 0
\label{a15}
\ee

\be
{\partial\over \partial z_+}\left(W(\phi)e^{-2\rho}F\right) =0
\label{b1}
\ee

\be
{\partial\over \partial z_-}\left(W(\phi)e^{-2\rho}F\right) =0
\label{b2}
\ee
{}From Eqs(\ref{b1},\ref{b2}) it follows that:
\be
F=q{e^{2\rho}\over W(\phi)}
\label{b3}
\ee
where $q$ is a constant.
\par
Substituting the above solution for $F$ into (\ref{a12}-\ref{a14}) and then
comparing the result
with Eqs(16-19) of Ref.\cite{domingo2}, we see that they are identical
providing we replace
\be
V(\phi)\to V(\phi) - {Gq^2l^2\over  W(\phi)} \,\,.
\ee
The rest of the proof therefore proceeds exactly as in Ref.\cite{domingo2}. As
long as
\be
{\partial \phi\over \partial z_+}{\partial \phi\over \partial z_-}>0 \, ,
\ee
it is always possible to choose a local coordinate system in which $\partial
\phi/\partial t=0$, so
that the field equations reduce to:
\be
\half {d \phi \over d x} - {1 \over 4} {{j}(\phi)\over l^2} +{1\over4} G q^2
k(\phi)
= - {1 \over4} C
\label{a36}
\ee
\be
e^{2\rho} = \half {d \phi \over d x}={1 \over 4} \left(-C +{j(\phi)\over l^2}
-G q^2 k(\phi)
\right)
\label{a37}
\ee
where $C$ is another constant, and $j(\phi)$ and $k(\phi)$  are defined by:
\bea
{dj(\phi)\over d\phi}&=& V(\phi)\\
{dk(\phi)\over d\phi}&=& {1\over W(\phi)}
\eea
Note that the constants of integration that result from the implicit
definitions of $j$ and $k$ above
can without loss of generality be absorbed into $C$. Also, since $\phi$ is
independent of time, so
is $\rho$ (cf. \req{a37}) and $F$ (cf.
\req{b3}). Thus, we have shown that (up to spacetime diffeomorphisms) every
solution is static,
and depends on two independent parameters: $q$ and $C$.
The solutions can be expressed in terms of  the spatial coordinate, $x$,
by inverting \req{a36}:
\be
x=x(\phi) = -2 \int {d\phi \over (C -{j(\phi)\over l^2} +   q^2 G k(\phi))}
\ee
The constant of integration here corresponds to a trivial shift $x\to x +
constant$.
The metric and electromagmetic field strength then take the simple form
\bea
2g_{+-}&=&e^{2\rho} =-{(C- {j(\phi)\over l^2} +  G q^2  k(\phi)) \over 4}\\
F&=& -q{ (C -{j(\phi)\over l^2} + q^2 G k(\phi))\over 4 W(\phi)}
\eea
\par
It is worth noting that in the case
${\partial \phi\over \partial z_+}{\partial \phi\over \partial z_-}<0$, one can
repeat the argument
above to show that there always exists a local coordinate system in which the
solution is
independent of the spatial coordinate\cite{domingo2}.
In this case, the solution is identical in form to the one given above, but
with the $x$-dependence
replaced by $t$-dependence.
\par
In order to verify that the constants of integration $q$ and $C$ are indeed
independent of
coordinate system and choice of $U(1)$ gauge we observe that for a
generic solution they can be written in the following invariant form:
\bea
 C&=&- g^{\mu\nu} \phi_{;\mu} \phi_{;\nu} +{j(\phi) \over l^2}-Gq^2 k(\phi)
\label{eq: covariant C}\\
q&=& {1\over \sqrt{-g}}  W(\phi) F  \,\, .
\label{eq: covariant q}
\eea
Clearly $q$ plays the role of
electric charge in the theory, while as shown below, $C$ determines the energy
of the solution.
\par
The above solutions  have a Killing vector which (for $e^{2\rho}>0$) is
timelike or spacelike depending on the sign of
$${\partial \phi \over \partial z_+}{\partial \phi \over \partial z_-}.$$
In fact one can verify  that Lie derivation along  the
vector field:
\be
k^\mu= {l\over \sqrt{-g}} \epsilon^{\mu\nu} \partial_\nu \phi
\label{eq: killing vector}
\ee
leaves the metric, the scalar field and the electromagnetic field strength
invariant  as long as the field equations (\ref{eq: field1}-\ref{eq: field3})
are satisfied. Thus the
Killing vector \req{eq: killing vector} represents a symmetry of every solution
in the generic
theory.
The norm of the Killing vector is:
\bea
|k|^2 &=& -l^2g^{\alpha\beta} \partial_\alpha \phi \partial_\beta \phi\none
  &=& l^2C- j(\phi) + l^2Gq^2k(\phi)
\label{eq: killing norm}
\eea
where we have used \req{eq: covariant C} to obtain the second line of the
above.
Defining
\bea
f(\phi; C,q) &=& - |k|^2\none
  &=& -(l^2C-j(\phi) + l^2 Gq^2 k(\phi))   \,\,\,  ,
\label{eq: defn f}
\eea
the necessary condition for  a given theory to admit
charged black hole configurations is the existence of
curves in spacetime given by $\phi(x,t) = \phi_0 = constant$ such that
\be
f(\phi_0; C,q):= j(\phi_0) -l^2 q^2 G k(\phi_0)-l^2C =0
\label{eq: horizon}
\ee
Note that for fixed $C$ there may exist critical values $q(C)$ for which
\be
\left.{df(\phi; C,q)\over \phi}\right|_{\phi_0}=0 .
\label{eq: extreme}
\ee
Thus
$\phi_0$ may be a local extremum of the function $f(\phi; C,q)$, or a point of
inflection. If it is an
extremum,  the norm of the Killing vector does not change sign as one moves
through the event
horizon at $\phi=\phi_0$. Moreover, if $q$ is varied away from its critical
value, the horizon in
general will either disappear, or two event horizons (an inner and outer
horizon) will appear.
In this case, the condition \req{eq: extreme} signals the presence of an
extremal black hole.
In the special case that one has a point of inflection, the norm of the Killing
vector does change
sign, but as the parameters are varied away from their critical values, one
expects the formation of
either one, or three horizons. These model independent conditions for the
presence of extremal
horizons will be useful in Section 5, when the thermodynamic properties are
discussed.
\par
We close this section on the space of solutions by writing down the most
general solution in a
particular convenient gauge. We choose the $x$-coordinate to be
$x=l\phi$, and $g_{tx}=0$. In this case, the metric takes the
``Reissner-Nordstrom" form:
\bea
ds^2 &=&  -f({x\over l}; C,q) dt^2 + f({x\over l}; C,q)^{-1}dx^2
\label{eq: RN}\\
F&=&{q\over W({x\over l})}
\eea
where $f$ is defined in \req{eq: defn f}. As we shall see in Section 5, this
form of the solution allows one to extract in a very simple way
the thermodynamic properties of the black hole. First, however, we will do a
Hamiltonian analysis
of the theory to prove that $C$ and $q$ are indeed a complete set of physical
configuration space
variables, and that the parameter $C$ determines  the energy of the solution.
\section{Hamiltonian Analysis}\medskip
\par
Spacetime is assumed to be locally a direct product $R\times \Sigma$, where the
spatial manifold,
$\Sigma$, can at this stage be either open or closed. The metric can be
parametrized as follows:
\be
ds^2=e^{2\rho}\left[-\sigma^2dt^2+\left(dx+Mdt\right)
^2\right]
.\label{eq:adm}
\ee
where $x$ is a local coordinate for the spatial section $\Sigma$ and $\rho$,
$\sigma$ and $M$ are
functions of spacetime coordinates $(x,t)$. In terms of this parametrization,
the action \req{eq: action 2} takes the form (up to surface terms):
\bea
S &=&\int dt dx[ {1\over G} ({\dot{\phi}\over \sigma}
(M\rho' + M' - \dot{\rho}) + {\phi'\over \sigma} (  \sigma\sigma' - MM' + M
\dot{\rho}
 + \sigma^2\rho' - M^2\rho')\nonumber\\
& &+\half\sigma e^{2\rho} {V(\phi)\over l^2}) + \half {e^{-2\rho}\over \sigma}
W(\phi) (\dot{A}_1- A'_0)^2]
\label{eq: action 3}
\eea
In the above dots and primes denote differentiation with respect to time and
space, respectively.
The canonical momenta for the fields
 $\{\phi, \rho, A_1\}$ are:
\bea
\Pi_\phi & = &{1\over G\sigma} (M\rho'+M'-\dot{\rho})
\label{eq: Pi phi}\\
\Pi_\rho &=& {1\over G\sigma} (-\dot{\phi} +M\phi')\\
\Pi_{A_1} & =& {e^{-2\rho}\over\sigma} W(\phi) (\dot{A}_1-A'_0)
\eea
The momenta conjugate to $\sigma$, $M$ and $A_0$ vanish: these fields play the
role of Lagrange
multipliers that are needed to enforce the first class constraints associated
with diffeomorphism
and gauge invariance of the classical action. A straightforward calculation
leads to the canonical
Hamiltonian (up to surface terms which will be discussed below):
\be
H_c= \int dx(M\fcal+ {\sigma\over 2G} \gcal + A_0 \jcal)
\label{eq: canonical hamiltonian}
\ee
where
\bea
\fcal&=& \rho'\Pi_\rho +\phi'\Pi_\phi-\Pi'_\rho \sim 0
\label{eq: f constraint}\\
\gcal&=& 2\phi''-2\phi'\rho' -2G^2\Pi_\phi \Pi_\rho - e^{2\rho} {V(\phi)\over
l^2} +G{e^{2\rho}\over
W(\phi)} \Pi^2_{A_1}\sim 0
\label{eq: g constraint}\\
\jcal &=& - \Pi'_{A_1} \sim 0
\label{eq: Gauss constraint}
\eea
are secondary constraints. The notation ``$\sim0$ " denotes weakly vanishing in
the Dirac
sense\cite{dirac}. It is straightforward to verify that the above constraints
are first class and that
no further constraints appear in the Dirac algorithm. The constraints $\fcal$
and $\gcal$ generate
spacetime diffeomorphisms.   The remaining constraint $\jcal$ is the analogue
of Gauss's law, and
generates U(1) gauge transformations. Since there are three first class
constraints and six phase
space degrees of freedom at each point in space, there are no propagating modes
in the theory. In
fact, the reduced phase space is finite dimensional. As in the vacuum
case\cite{domingo1}, the
constraints can be explicitly solved for the momenta as a function of the
fields:
\bea
\Pi_\rho &=&Q[C, q; \rho, \phi]
\label{eq: p1}\\
\Pi_\phi &=& { g[q; \rho, \phi]\over (2G)^2 Q[C, q; \rho, \phi]}
\label{eq: p2}\\
\Pi_{A_1} &=& q
\label{eq: p3}
\eea
where we have defined:
\bea
Q[C, q; \rho, \phi]&\equiv& {1\over G} \sqrt{(\phi')^2 + e^{2\rho}(C -
{j(\phi)\over l^2} +Gq^2  k(\phi))}
\label{eq: Q definition}\\
g[q; \rho, \phi] &\equiv& 4\phi'' - 4\phi'\rho' -2 e^{2\rho}{V(\phi)\over l^2}
   + 2G q^2 {e^{2\rho}\over W(\phi)}
\label{eq: g definition}
\eea
In the above, $C$ and $q$ are arbitrary constants of integration which
represent physical
observables in the theory. Using  \req{eq: p1} and \req{eq: p3}, these
observables can be written
in terms of the original phase space variables as follows:
\be
C = e^{-2\rho}(G^2 \Pi_\rho^2 -(\phi')^2) +{j(\phi)\over l^2} -G k(\phi)
\Pi_{A_1}^2
\label{eq: C}
\ee
\be
q= \Pi_{A_1}
\label{eq: q}
\ee
It is straightforward to show that  $C$ and $q$ as defined above have vanishing
Poisson brackets
with all the
constraints: they are physical observables in the Dirac sense. Moreover,
both $C$ and $q$ are independent of the spatial coordinate on the constraint
surface. $q'\sim 0$
follows trivially from Gauss' law, while from \req{eq: C},
\req{eq: q} and (\ref{eq: f constraint} - \ref{eq: Gauss constraint}) one can
show that:
\be
\gtilde:=C' = -e^{-2\rho}(\phi'\gcal +2G^2\Pi_\rho\fcal) +2G k(\phi)\Pi_{A_1}
\jcal
\sim 0
\label{eq: C'}
\ee
Since the observables $C$ and $q$ have vanishing Poisson bracket with each
other,
the reduced phase space is four dimensional. The momenta conjugate to $C$ and
$q$
are\footnote{In practice, these were calculated by differentiation of the exact
Hamilton-Jacobi
functional derived in Section 6 with respect to the observables $C$ and $q$.}:
\bea
p_C= - \half\int dx {e^{2\rho} \Pi_\rho\over ((G \Pi_\rho)^2 - (\phi')^2)}
\label{eq: pc}\\
p_q= -\int dx \left(A_1 + {G k(\phi) e^{2\rho} \Pi_\rho \Pi_{A_1} \over
   ((G \Pi_\rho)^2 - (\phi')^2)}\right)
\label{eq: pq}
\eea
It is straightforward to verify the canonical Poisson bracket relations:
\bea
\{C, p_C\} = \{ q, p_q\} =1\\
\{C,q\}= \{C,p_q\} = \{ q, p_C\} = \{p_C,p_q\} =0
\eea
As in the case of pure dilaton gravity\cite{domingo1}, the apparent discrepancy
between the
generalized Birkhoff theorem (which states the existence of a two parameter
family of solutions up
to spacetime diffeomorphisms and gauge transformations) and the dimension of
the reduced phase
space is explained by the following observation. The configuration space
variables $C$ and $q$ are invariant under a general canonical transformation
generated by:
$$ \int dx \left(\xi \fcal + \eta \gcal + \chi \jcal\right) \, ,$$
 independent of the
boundary conditions on the test functions. The momenta, on the other hand,
transform as
follows:
\bea
\delta p_C &=& -\half\int dx \left( { e^{2\rho}(\xi\Pi_\rho - 2\eta \phi')\over
 ((G\Pi_\rho)^2 - (\phi')^2) } \right)'\\
\delta p_q &=& -\int dx \left( {G k(\phi) e^{2\rho} \Pi_{A_1} \over
  ((G\Pi_\rho)^2 - (\phi')^2) }  (\xi \Pi_\rho -2\eta \phi')+\chi\right)'
  \eea
For non-compact spacetimes, the momenta are invariant  only if the
test functions vanish sufficiently rapidly at infinity. Since symmetry
transformations in the
canonical theory are defined precisely in terms of test functions that vanish
at infinity, $p_C$ and
$p_q$ are physical observables, as claimed above. However, in the proof of
Birkhoff's theorem,
the global behaviour of the spacetime diffeomorphisms and gauge transformations
is not restricted,
and hence the momenta are not invariant in the covariant sense:
both $p_C$ and $p_q$ can be changed by ``non-canonical", large diffeomorphisms
and/or U(1)
gauge transformations.  As shown for SSG by Kuchar\cite{ssg}, and for generic
2-D
dilaton
gravity\cite{gkm}, the momentum conjugate to $C$ has a physical interpretation
as the time
separation at infinity. $p_q$, on the other hand, is clearly related to the
asymptotic choice of
$U(1)$ gauge.
\par
 We also note that the transformation properties of $p_C$ and $p_q$  under
large diffeomorphisms and U(1) gauge transformations play an important role in
the quantum theory. In particular, they ensure that physical quantum wave
functions  have the right transformation properties under such large
transformations. For example, states of definite charge $q$ should pick up a
phase proportional to
$q\delta p_q$. This will be discussed further in the Section on Dirac
quantization.
\par
We now address the question of surface terms and derive the ADM energy of the
generic solution.
The analysis is simplified by first using \req{eq: C'} and \req{eq: Pi phi} to
write the canonical
Hamiltonian  in the following form:
\be
H_c =  \int dx \left[  {\dot{\phi}\over\phi}\fcal -{1\over 2G}\left({\sigma
e^{2\rho}
\over  \phi'}\right)\gtilde -A_0 q' \right]
\label{eq: hamiltonian 2}
\ee
For configurations that approach \req{eq: RN}
asymptotically, so that $\dot{\phi}\to0$ and
\be
{\sigma e^{2\rho}\over l\phi'}\to 1
\ee
at spatial infinity, the surface term that is required in order to ensure that
the canonical Hamiltonian
yields the correct equations of motion is:
\be
H_{surface} = H_{ADM} + \int (A_0 q)'
\ee
where we have defined the ADM surface Hamiltonian:
\be
H_{ADM} = {l\over 2G} \int dx \left( {\sigma e^{2\rho}\over l \phi'} C\right)'
\label{eq: ADM Hamiltonian}
\ee
The resulting ADM energy, which is by definition equal to $H_{ADM}$ evaluated
at spatial
infinity is:
\be
M= {lC\over 2G}
\label{eq: M}
\ee
Thus $C$ is indeed related to the energy of the solution, as claimed above.
Similarly, the
``electric charge" $q$ can be seen to be the conserved charge associated with
U(1) gauge
transformations
that leave $A_0$ invariant at spatial infinity.
Note that this derivation of the ADM energy does not require asymptotic
flatness\footnote{The
derivation of the ADM mass in generic dilaton gravity has recently been
discussed in \cite{kim}
and \cite{kummer}}, merely that
the solutions approach the static Reissner-Nordstrom form given in the previous
sections. This is
important since the black hole solutions are not asymptotically flat in the
generic case: for example
in J-T gravity, the solution is asymptotically deSitter.
\par
For completeness we now write down the Hamiltonian equations of motion that
follow from the
canonical Hamiltonian \req{eq: hamiltonian 2} when supplemented with the above
surface term:
\bea
\dot{\Pi}_\phi &:=&\{\Pi_\phi, H_c\}\nonumber\\
   &=&  -{\sigma''\over G} - {1\over G} (\sigma\rho' )'
 +{\sigma\over 2G} {e^{2\rho}\over l^2} {dV\over d\phi} + (M\Pi_\phi)' +
{\sigma e^{2\rho}\over 2W^2}{dW\over d\phi} \Pi^2_{A_1}\\
\dot{\Pi}_\rho &:=& \{\Pi_\rho, H_c\} \nonumber\\
    &=&(M\Pi_\rho)' -{1\over G} (\sigma\phi')'+{\sigma\over 2G} {e^{2\rho}\over
l^2} V(\phi)
-{\sigma e^{2\rho}\over W(\phi)} \Pi^2_{A_1}\\
\dot{\Pi}_{A_1} &=& \{\phi_{A_1}, H_c\}\none
   &=& 0
\eea
The physical observables $C$ and $q$ commute with the canonical Hamiltonian,
and are therefore
time independent, as expected.
\medskip
\section{Thermodynamic Properties}
The surface gravity,  $\kappa$ of a black hole  is given in terms of the
Killing vector by\cite{wald2}:
\be
\kappa^2 = -\left.\half \nabla^\mu k^\nu\nabla_\mu k_\nu\right|_{\phi_0}
\ee
 $\phi=\phi_0$ is a solution  \req{eq: horizon}.  Using the expression \req{eq:
killing vector} for
the
Killing vector  and the field equation \req{eq: field2}
one obtains (after some tedious algebra):
\be
\kappa = {V(\phi_0)\over 2l}- {l q^2 G \over 2 W(\phi_0)}
\label{eq: kappa}
\ee
\par
The Hawking temperature in the generic theory  can be calculated by
analytically continuing the
solutions \req{eq: RN} to Euclidean spacetime and imposing periodicity in the
imaginary time
direction
in order to allow non-singular solutions
without boundary at the outer horizon. The Euclidean solutions take  the
general form:
\be
ds^2 = f({x\over l}; M, q) dt_E^2 + {1\over f({x\over l}; M, q) }dx^2
\label{eq: RN2}
\ee
where $M= lC/2G$, and
$
f({x\over l}; M, q) $
is the square of the norm of the Killing vector defined in \req{eq: defn f}. By
defining a
dimensionless time coordinate $\alpha:= t_E/ a$ and a
new radial coordinate, $R$, such that
\be
R^2 = a^2 f({x\over l};M,q)
\ee
the metric can be placed in the form:
\be
ds^2 = R^2 d\alpha^2 + H(R; M,q) dR^2
\ee
where
\be
H(R; q, M) := {4\over a^2 (f'({x\over l}; M,q))^2}
\ee
and the prime denotes differentiation with respect to $x$.
The solution will be regular at  $R=0$, if
$H(R) \to 1$ as $R\to 0$ and the coordinate $\alpha$ be periodic with period
$2\pi$.  The
Hawking temperature is then the inverse of the period of the corresponding
Euclidean time
coordinate:
\be
T = \left( 2\pi a\right)^{-1} = { f'(x_0/l;M,q)\over 4\pi}
\label{eq: hawking temp}
\ee
where $x_0/l=\phi_0$ is the location of the horizon. Using  \req{eq: defn f}
and the fact that  $j' = V/l$ and $k' = 1/(lW)$, we are able to verify the
standard relationship between the Hawking temperature and the surface gravity
in the generic
theory:
\be
\kappa = 2\pi T
\ee
Since \req{eq: extreme} signals the presence of an extremal black hole
at $\phi=\phi_0$, this shows that the surface gravity and Hawking temperature
both vanish for extremal
black holes in the generic theory.
\par
{}From the form of \req{eq: RN2} and \req{eq: defn f} it is also possible to
extract directly the
entropy of the black hole.  In particular, if we vary the solution
infinitesmally, but stay on the event
horizon $\phi_0$ which is a solution to $f(\phi_0; M,q) = 0$, we obtain the
condition:
\bea
0 &=& \delta f(\phi_0; M, q)\none
   &=& -2Gl\delta M + 2l\kappa\delta \phi - 2ql^2 G k(\phi_0) \delta q
\label{eq: first law 1}
\eea
where $\kappa$ is precisely the surface gravity defined in \req{eq: kappa}.
This gives:
\be
\delta M = \left({V(\phi_0)\over l} - {l q^2 G\over W(\phi_0)}\right) {\delta
\phi_0\over 2G} - lq k(\phi_0)
\delta q
\ee
which is of the desired form:
\be
\delta M = \kappa \delta S - {\cal{P}} dq\ee
with entropy
\be
S = {2\pi\over G} \phi_0
\label{eq: entropy}
\ee
and generalized force ${\cal P}:= qlk(\phi_0)$  associated with the charge $q$.
Note that the
entropy \req{eq:
entropy} is proportional to  the value of the scalar field at the event
horizon. \req{eq: entropy} has precisely the
same form as the expression obtained for general dilaton gravity without $U(1)$
gauge field in \cite{gkm}.
The above analysis implies that in dilaton gravity, \req{eq: horizon} can be
interpreted as the thermodynamic equation of state for the black
hole since it relates the energy $M$ and entropy ${2\pi\phi_0\over G}$ to the
extrinsic macroscopic variable
$q$.
\par
Although \req{eq: entropy} was obtained using a convenient choice of
coordinates, it is in fact
coordinate invariant. We will now verify this by  using  Wald's general
method\cite{wald} to
derive the entropy of black hole solutions in the generic model. The
application of the method in
1+1 dimensions turns out to be simpler than in higher dimensions. Here we use
the language of
tensor calculus but the analysis can easily be repeated using forms as done by
Wald. In general one
looks at the variation of the action under space-time diffeomorphisms of the
form:
\bea
x^\mu\to x^{'\mu}&=&x^\mu+\delta x^\mu\none
\Phi^A(x)\to \Phi'{}^A(x') &=& \Phi^A(x) + \delta \Phi^A(x)
\eea
where for the moment we use a condensed notation in which the complete set of
fields (including
the metric) is denoted by $\Phi^A(x)$, and $x$ is the spacetime coordinate. It
can be shown that,
under such a general  transformation,  an  action which is second order in
derivatives of the fields
has the following variation:
\be
\delta I = \int d^2x \left( {\delta I\over \delta \Phi^A} \delta \Phi^A
+ {\partial j^\mu \over \partial x^\mu}\right)
\label{eq: delta I}
\ee
where $j^\mu$ is the associated Noether current.
Diffeomorphism invariance of the action requires that the Noether current be
divergence free when
the classical field equations
\be
{\delta I\over \delta \Phi^A}=0
\ee
are satisfied.
\par
For the action \req{eq: action 2} the Noether current is:
\bea
j^\lambda &=& \left( {1\over 2G} (\phi R + {V\over l^2}) - {1\over 4} W
F^2\right) \delta x^\lambda   -{1\over 2G} \nabla_\sigma\phi (
g^{\alpha\lambda} g^{\beta\sigma} - g^{\alpha\beta}
g^{\lambda\sigma} ) \deltabar g_{\alpha\beta}\none
 & & + {1\over 2G} \phi( g^{\alpha\sigma} g^{\beta \lambda} - g^{\alpha\beta}
g^{\sigma\lambda}
) \nabla_\sigma(\deltabar g_{\alpha\beta})
- W F^{\lambda\sigma }\deltabar A_\sigma
\label{eq: noether 1}
\eea
where $\deltabar$ denotes variation of the corresponding field under Lie
derivation along $\delta x^\mu$.
As shown in Section 2, the solutions have a single Killing vector, so we take:
\be
\delta x^\lambda = - k^\lambda = -{l\over \sqrt{-g}} \epsilon^{\lambda\sigma}
\nabla_\sigma \phi
\ee
The variations of the scalar field and metric vanish for such transformations,
but the Lie derivative
of the vector potential in the Killing direction can be
written:
\be
\deltabar A_\sigma= -k^\eta F_{\sigma\eta} + \nabla_\sigma(k^\eta A_\eta)
\ee
Using this expression and the field equations, we write the Noether current as:
\bea
j^\lambda &=&- \left[ {1\over 2G} \left(-{\phi\over l^2}  {dV \over d\phi} +
{V\over l^2}\right) + {1\over 4} (\phi{dW\over d\phi}
- W) F^2 \right]
{l \over  \sqrt{-g} } \epsilon^{\lambda\sigma} \nabla_\sigma\phi\none
& &+ W F^{\lambda\sigma} F_{\sigma\eta}
 {l\over \sqrt{-g} }\epsilon^{\eta\rho} \nabla_\rho\phi
   -\nabla_\sigma\left({l\over \sqrt{-g}}
 W F^{\sigma\lambda}     \epsilon^{\eta\rho} \nabla_\rho \phi A_\eta\right)
\label{eq: noether 2}
\eea
Note that the last term is gauge dependent, but  has identically vanishing
divergence. The Noether current is not uniquely defined in general since one
can always add to it
an arbitrary divergence free term. As shown by
Wald, however there is  a unique diffeomorphism covariant current which
contains at most first
derivatives of the fields. In
the present case we must also add the condition of gauge invariance, in which
case, we find the
following, unique Noether current:
\be
j^\lambda= - \left[{1\over 2G} \left(- {\phi\over l^2} {d V\over d\phi} +
{V\over l^2}\right) - \half
  q^2 ( {\phi \over W^2} {dW\over d\phi} + {1\over W})\right]
 {1\over \sqrt{-g}} \epsilon^{\lambda\sigma}\nabla_\sigma \phi
\label{eq: final j}
\ee
Following Wald, we can define the associated one form:
\bea
J_\mu &:=&\sqrt{-g} \epsilon_{\mu\lambda} j^\lambda\\
 &=& - \left[{1\over 2G l} \left( V -\phi {d V\over d\phi}\right) - \half
  q^2 l\left( {\phi \over W^2} {dW\over d\phi} + {1\over W}\right)\right]
  \nabla_\mu\phi
\label{eq: def J}
\eea
Since the exterior derivative of $J$ vanishes identically we can at least
locally write it as the derivative of a zero form:
\be
J_\mu = {\partial Q\over \partial x^\mu}
\ee
where
\be
Q =  {1\over 2G l} ( V - { G l^2 q^2\over W} ) \phi  - {1\over G}
 g^{\alpha\beta} \nabla_\alpha\phi\nabla_\beta \phi  \,\,.
\ee
According to Wald's prescription, the value of $Q$ at the event horizon
should be proportional to the entropy. Indeed, since
on the event horizon $|\nabla \phi|^2=0$, we find:
\bea
Q|_{horizon} &=&  \left.{1\over 2Gl} ( V- {G l^2q^2 \over W})\phi
\right|_{horizon}
  \\
   &=& {\kappa\over 2\pi}S
\eea
where $\kappa$ is the surface gravity and $S$ is  the entropy defined in
\req{eq:
entropy} above.
  \medskip
\section{Dirac Quantization}
\par
We will now quantize the generic theory in the functional Schrodinger
representation, using
techniques first developed by Henneaux\cite{henneaux} to quantize
Jackiw-Teitelboim gravity,
and later extended to the generic dilaton gravity theory in \cite{domingo1}.
Similar techniques
have recently been applied by Strobl\cite{strobl} to the quantization of
dilaton gravity coupled to
a Yang-Mills field in a first order formalism. For the pure dilaton-gravity
sector of String Inspired Dilaton gravity, these techniques have recently been
shown\cite{benedict} to yield quantum theories equivalent to those obtained
within the gauge theoretical formulation\cite{cangemi}.
\par
Following the standard Dirac\cite{dirac} prescription we first define a Hilbert
space of
states described by
wave functionals $\psi[\rho,\phi, A_1]$ defined on the unreduced configuration
space. Given a
Hilbert space scalar product of the form:
\be
<\psi|\psi> = \int \prod_x d\rho(x) d\phi(x) dA_1(x)
  \mu[\rho,\phi,A_1] \psi^* \psi
\label{eq: measure}
\ee
we can define Hermitian operators for the momenta canonically conjugate to
$\rho$, $\phi$ and
$A_1$ as follows:
\bea
\hat{\Pi}_\rho &=&-i\hbar {\delta\over \delta \rho(x)} + {i\hbar\over 2}
    {\delta  \ln (\mu[\alpha,\tau])\over \delta\rho }\\
\hat{\Pi}_\phi &=&-i\hbar {\delta\over \delta \phi(x)} + {i\hbar\over 2}
     {\delta \ln (\mu[\alpha,\tau])\over\delta \phi}\\
\pa &=& -i\hbar {\delta\over \delta A_1(x)} + {i\hbar\over 2}
     {\delta \ln (\mu[\alpha,\tau])\over\delta A_1}
\label{eq: momenta}
\eea
\par
Physical quantum wave functionals $\Psi[\rho,\phi,A_1]$ are defined to be those
states annihilated
by the quantized constraints:
\bea
\hat{\fcal} \Psi = 0\\
\hat{\gtilde}  \Psi = 0\\
\hat{\jcal} \Psi = 0
\eea
The first and last constraints are linear in the momenta and the standard
factor ordering  with all
the momenta on the right is self-adjoint providing that the functional measure
$\mu$ is invariant
under the corresponding gauge transformation. The factor ordering for the
Hamiltonian constraint
$\gtilde$  on the other hand is more subtle. Here we will be concerned only
with the lowest order,
or WKB approximation, and hence we can safely defer questions concerning the
choice of
factor ordering and measure.
\par
To proceed we look for the analogue of stationary states in the theory. In
particular, since the
observable $C$ defined in \req{eq: C} determines the energy of a configuration,
we
look for physical states that are also eigenstates of the operator $\hat C$
\be
\hat{C} \Psi = C\Psi
\label{eq: c eigenstate}
\ee
where $C$ is a constant.
Note  the \req{eq: c eigenstate} automatically guarantees that the Hamiltonian
constraint  is also
satisfied, since:
\be
\hat{\gtilde} \Psi =C' \Psi = 0
\ee
Analogously, the $U(1)$ gauge constraint can be satisfied by looking for
eigenstates of the charge
operator:
\be
\hat{q}\Psi = \pa \Psi = q \Psi
\ee
Thus we are looking for  (spatial) diffeomorphism invariant energy and charge
eigenstates. Since $C$ and $q$ are a complete set of variables, one can
construct all states by
taking  linear combinations of such states.
\par
The WKB approximation for gauge
theories in general and quantum gravity in particular has been discussed
extensively by
Barvinsky\cite{barvinsky} and more recently by  Lifschytz {\it
al}\cite{mathur}. We expand the
phase of the wave functional as follows:
\be
\psi[\rho,\phi,A_1] = exp {i\over\hbar}[S_0[\rho,\phi,A_1] + \hbar
S_1[\rho,\phi,A_1]
   +...]
\ee
To lowest order in $\hbar$ we find that $S_0$ must obey the following
equations:
\be
\left[e^{-2\rho}\left( G^2 \left({\delta S_0\over \delta \rho}\right)^2
- (\phi')^2\right) + {j(\phi)\over l^2} - G k(\phi)
\left({\delta S_0 \over \delta A_1}\right)^2
\right]= C
\label{eq: HJE}
\ee
\be
\rho' {\delta S_0 \over \delta \rho} + \phi' {\delta S_0\over \delta \phi}
- \left({\delta S_0 \over \delta \rho}\right)' = 0
\label{eq: diff 1}
\ee
\be
{\delta S_0 \over \delta A_1} = q
\label{eq: s1}
\ee
Note that \req{eq: HJE} is the Hamilton-Jacobi equation for the theory.
Remarkably, the above
functional differential equations can be solved exactly.  After some algebra,
Eqs.(\ref{eq: HJE}) and
(\ref{eq: diff 1}) reduced to the following first order functional differential

equations for $S_0$:
\bea
{\delta S_0 \over \delta \rho} &=& Q
\label{eq: s2}\\
{\delta S_0 \over \delta \phi} &=& {g\over (2G)^2 Q}
\label{eq: s3}
\eea
where $g$ and $Q$ are defined in Eqs.[\ref{eq: Q definition}] and [\ref{eq: g
definition}].
The closure of the constraint algebra guarantees that these functional
differential equations are integrable. The resulting exact solution for the
Hamilton principal functional,
 $S_0$, is:
\be
S_0[C,q;\rho,\phi,A_1] = \int dx \left[ Q +
 {\phi'\over 2G} \ln \left({\phi'-GQ\over \phi' +G Q}\right) + q A_1\right]
\label{eq: wave function}
\ee
As expected, the partial derivatives of $S_0$ with respect to $C$
and $q$ are proportional (on the constraint surface) to the corresponding
conjugate momenta, $p_C$ and
$p_q$, as given in equations
\req{eq: pc} and \req{eq: pq}. This ensures (at least to the WKB order
considered here) that the quantum operators $\hat{p}_C$ and
$\hat{p}_q$ have the correct action on eigenstates of $\hat{C}$ and $\hat{q}$.
i.e. $\hat{p}_C |C> = i\hbar (\partial/\partial C) |C>$, etc. Moreover,  it is
straightforward to verify that
under a  large $U(1)$  gauge transformation, the change in the WKB phase is
precisely equal to $q \delta p_q$ as expected from the discussion in Section 4.
The action of large spacetime diffeomorphisms is more complicated since it is
not generated by constraints linear and homogeneous in the momenta. However, as
long as the quantum operators $\hat{C}$ and $\hat{p}_C$ are correctly
represented, energy eigenstates will necessarily change by a phase under a
change in $p_C$, which classically is generated by large spacetime
diffeomorphisms.
\par
We also note that one can also interpret the WKB wave-functional $\Psi_0 =
\exp(i/\hbar S_0)$ in
a different way.  If one chooses to first solve the constraints in the theory
classically as in
Eqs.[\ref{eq:
p1}-\ref{eq: p3}]and then quantize, one obtains the complete set of quantum
constraints on
physical states:
\bea
\left({\prho} - Q\right)\Psi_0 &=& 0\\
\left({\pphi} - {g\over (2G)^2 Q}\right)\Psi_0 &=& 0\\
\left({\pa} - q\right)\Psi &=& 0
\eea
In this case, $\Psi_0:= \exp i S_0 /\hbar$ provides an exact solution to the
constraints, providing that the quantum
operators are defined to be Hermitian with respect to a trivial functional
measure $\mu=1$.
Thus, $\Psi_0$ can be interpreted \cite{henneaux},\cite{domingo1} as an exact
physical wave functional for the
theory in which the original constraints are replaced by their classical
solutions Eqs.[\ref{eq:
p1}-\ref{eq: p3}]. The quantum theories are equivalent classically and differ
by factor ordering.
\par
Finally we comment on the relationship between the analytic  structure in the
WKB
phase and the existence of event horizons. The logarithm in \req{eq: wave
function} appears to
develop a singularity when
\be
(\phi')^2 - G^2 Q^2 = 0
\ee
which, as can be seen from \req{eq: killing norm} and \req{eq: Q definition}
occurs at the event horizon ($|k|^2=0$) in any non-singular coordinate system
($e^{2\rho}\neq0$). Depending
on the nature of the
analytic continuation in defining the quantum wave function, the phase can pick
up an imaginary
contribution for values of $x$ for which $(\phi')^2 - G^2 Q^2 < 0$. As can be
seen from \req{eq: killing norm} and \req{eq: Q definition}, if one chooses
nonsingular coordinates along a spacelike slice (for which $e^{2\rho}>0$), this
corresponds to the regions in spacetime for which the
Killing
vector is spacelike. For example, for non-extremal black holes in the generic
theory it is possible to go to Kruskal-like coordinates $(X, t_K)$, in which
case:
\be
Im S_0 = {1\over 2G} Im \int dX {\partial \phi \over \partial X} \ln \left(
  {X-t_K\over X+t_k}\right)
\ee
If one chooses a spacelike slice that cuts across both horizons of an eternal
black hole this yields:
\be
Im S_0 = {i\pi \over 2G} (\phi_+ - \phi_-) = 0
\label{eq: im.s}
\ee
where $\phi_\pm$ corresponds to the value of the scalar field at the horizon
$X\pm t_K$ in the left and right asymptotic regions, respectively.
\par
 On the other hand, if one chooses a slicing that only cuts through
the right horizon, say, as is appropriate in the case where the event horizon
is formed by collapsing matter, then the imaginary part of the WKB phase is not
necessarily zero. It will have a term proportional to $\phi$ evaluated at the
horizon, i.e. the entropy of the black hole. This analysis suggests that there
is an intriguing connection between the entropy  of the black hole, and the
imaginary
part of its WKB phase. Such  a connection is in some sense natural, since
the imaginary part of the WKB phase is generally related to quantum mechanical
tunnelling, whereas Hawking radiation can heuristically be thought of as a
quantum mechanical
instability.  In light of this interpretation, \req{eq: im.s} is also
consistent with
the recent result of Martinez\cite{martinez} who used semi-classical partition
function techniques to prove that the entropy of an eternal black hole vanishes
due to a cancellation between contributions from the left and right hand
asymptotic regions of the Kruskal diagram. It is also important to note that in
the case of extremal black holes, the Killing vector is never spacelike, so
that the imaginary part of the
WKB phase is identically zero. This result  is consistent with recent
work\cite{extreme2} which argues that the entropy of extreme black holes is
zero. It is also consistent with the analysis of
Kraus and Wilczeck\cite{kraus}, who showed
semi-classically that gravitational back reaction corrections yield a vanishing
amplitude for radiative
processes at extremality.
\par\indent
\section{Examples}
\subsection{Spherically Symmetric Gravity}
We start with the Einstein-Maxwell action in 3+1 dimensions:
\be
I^{(4)}= {1\over 16\pi G^{(4)}}\int d^4x \sqrt{-g^{(4)}} \left(R(g^{(4)})
-(F^{(4)})^2\right)
\ee
We impose spherical symmetry on the electromagnetic field, and on the metric,
as follows:
\be
ds^2=g_{\mu\nu}(x)dx^\mu dx^\nu + {l^2\phi^2\over 2}d\Omega^2,
\ee
where the $x^\mu$ are coordinates on a two dimensional spacetime $M_2$
with metric $g_{\mu\nu}(x)$ and $d\Omega^2$ is the line element of the 2-sphere
with area $4\pi$.  The corresponding spherically symmetric solutions to the
Einstein Maxwell
equations
are the stationary points of the dimensionally reduced action
functional\cite{thomi}:
\be
\tilde I[g,\phi]=\int_{M_2} d^2x \sqrt{-g}\left[ \half\left({\phi^2\over
4}R(g)+\half g^{\mu\nu}
\partial_\mu\phi\partial_\nu\phi +{1\over l^2}\right) - {1\over 8}
\phi^2F^{\mu\nu}F_{\mu\nu} \right],
\ee
where  $l^2= G^{(4)}$ is the square of the 3+1 dimensional Planck length.
This action is of the same form as \req{eq: action 1} above. It can be verified
that after the appropriate
reparametrization as described in Section 2, the reduced action takes the form
of \req{eq: action
2}, with
 the corresponding values for the potential $V(\phi)$ and the function
$W(\phi)$ are
respectively\footnote{In this section we set $G=1$ for simplicity}:
\be
V(\phi) = {1\over \sqrt{2\phi}}
\ee
\be
W(\phi) = (2\phi)^{3/2}
\ee
The most general solution to the field equations in our parametrization can be
written in the form:
\bea
ds^2&=& - {r\over l}\left(1-{2M l^2\over r} + {l^4 Q^2\over r^2}\right)dt^2 +
{r\over l} {dr^2\over \left(1-{2Ml^2\over r} + {l^4Q^2\over r^2}\right)}\\
\phi &=& \half {r^2\over l^2} \\
F&=& {l^2Q\over r^2}
\eea
The physical observables \req{eq: C} and \req{eq: q}
are $C=2M/l$ and $q=Q$, as expected. Note that the above metric corresponds to
the usual Reissner-Nordstrom metric up to the conformal reparametrization:
\be
g_{\mu\nu}= \sqrt{2\phi}g_{\mu\nu}^{RN}
\ee
As is well known, there are two event horizons, located at $r_\pm=
l^2(M\pm\sqrt{M^2-Q^2})$. From the expressions
\req{eq: kappa} and \req{eq: entropy} we can calculate the standard
expressions\cite{wald2} for
the  surface gravity, and entropy, respectively, for this model:
\bea
\kappa&=& \sqrt{M^2-Q^2}\over l^2(M+ \sqrt{M^2 -Q^2})^2\\
S&=& {A_+\over 4 l^2}
\eea
where $A_+$ is the area of the outer horizon:
\be
A_+ := 4\pi l^2(M+ \sqrt{ M^2 -Q^2})^2
\ee
The first law then takes the expected form:
\be
\delta M = {\kappa\over 2\pi} \delta S + \varphi \delta q
\ee
where $\varphi$ is the electrostatic potential at the horizon.
\subsection{Achucarro-Ortiz Black Hole}
\par
This black hole is a solution to the field equations for the Jackiw-Teitelboim
theory of gravity,
which can be obtained by imposing axial symmetry in 2+1 gravity\cite{ortiz}. It
is therefore the projection of the BTZ black hole\cite{BTZ}. In particular, one
starts with the Einstein action in $2+1$ dimensions, with cosmological constant
$\Lambda$:
\be
I^{(3)} = \int d^3x \sqrt{-g^{(3)}}(R^{(3)} + \Lambda)
\ee
and then restricts consideration to the most general axially symmetric metrics:
\be
ds^2 = g_{\mu\nu}dx^\mu dx^\nu + \phi^2(x)[d\theta + A_\mu(x) dx^\mu]^2
\ee
where in this case $x^\mu = \{t,\rho\}$ refer to cylindrical coordinates in the
2+1 spacetime, with angular coordinate $\theta$.
One then finds a reduced two-dimensional action of the form:
\be
I = \int d^2x \sqrt{-g} \phi\left( R + \Lambda -{1\over 4} \phi^3 F^{\mu\nu}
F_{\mu\nu}\right) .
\ee
In the above, $F_{\mu\nu}$ is the field strength of the one form $A_\mu$ that
appears in the
parametrization of the axially symmetric metric above.  If we define $l^2:=
1/\Lambda$, then this is again of the desired form with $V=
\phi$ and
$W= \phi^3$. The most general solution corresponds precisely to the dimensional
reduction of the
BTZ axially symmetric black hole:
\bea
\phi&=& \sqrt{\Lambda}r\\
ds^2&=& - ({\Lambda^2 r^2\over 2} - \sqrt{\Lambda}M + {J^2\over 4\Lambda r^2})
dt^2 + {dr^2\over ({\Lambda^2 r^2\over 2}
-\sqrt{\Lambda}M + {J\over 4\Lambda r^2})} \\
F&=& {1\over \Lambda^{3/4}}{J\over r^3}
\eea
In this case, the phase space observables are $C=M$ and $q=-J$. Note that the
$U(1)$ charge is related to the angular momentum of the BTZ black hole. There
are again two event
horizons, with locations:
\be
r_{\pm} = {1\over \Lambda^{3/4}}\left({M\pm\sqrt{M^2- J^2/2}}\right)^\half
\ee
One can again compute the surface gravity and entropy associated with the outer
horizon from the
expressions given above:
\bea
k&=& {\Lambda\over2} {(r_+^2-r_-^2)\over r_+}\\
S &=&4\pi \sqrt{\Lambda}r_+
\eea
\par\noindent
\subsection{String Inspired Model}
\par
The string inspired model with an electromagnetic interaction was studied in
detail in \cite{frolov}. In our
parametrization, the string inspired action (without tachyon fields) is of the
form:
\be
S= \half \int d^2x \sqrt{-g} \left[\phi R + \lambda^2
-\half \phi^2 g^{\alpha\beta} g^{\gamma\delta} F_{\alpha\gamma} F_{\beta\delta}
\right]
\label{eq: sig action}
\ee
Thus, if  $l^2:= 1/\lambda^2$, then   $V(\phi) = 1$ and $W(\phi) = \phi^2$. The
most general solution in
this parametrization can be written\cite{frolov}:
\bea
ds^2 &=& -\left(\lambda r -{2M\over\lambda} + {Q^2\over \lambda^3 r}\right)
dt^2
+
 \left(\lambda r -{2M\over \lambda} + {Q^2\over \lambda^3 r}\right)^{-1} dx^2
\\
\phi &=& \lambda r\\
F&=& {Q\over \lambda^2 r^2}
\eea
These solutions  describe black holes with mass $M$ and charge $Q$. The event
horizons
occur at
\be
r_\pm := {M\over \lambda^2} \pm {1\over \lambda^2}  \sqrt{M^2 -Q^2}
\ee
Using formulae \req{eq: kappa} and \req{eq: entropy}, respectively, we obtain
the surface gravity
and entropy:
\bea
\kappa&=& \half \left(\lambda - {Q^2\over \lambda^3 r_+^2}\right) =
{ \lambda\sqrt{M^2 - Q^2}\over M+ \sqrt{M^2-  Q^2}}\\
S&=& 2\pi r_+ = {2\pi\over \lambda} (M+ \sqrt{M^2- Q^2})
\eea
These expressions coincide with the ones found in Ref.\cite{frolov}, and one
can verify that the first law:
\be
\delta M = {k\over 2\pi} \delta S + {Q\over r_+} \delta Q
\ee
is obeyed.
\par\noindent
\section{Conclusions}
We have analyzed in detail the space of solutions,  physical phase space,
classical
thermodynamics and quantum mechanics of the most general dilaton gravity theory
coupled to a $U(1)$ gauge
potential. It was shown that the surface gravity, Hawking
temperature and the imaginary part of the WKB phase all vanish for extremal
black holes.
 In a future publication\cite{sutton} we will
examine in more detail the properties of the WKB wave functionals in the
generic theory, both with and without $U(1)$ charge. We also hope to extend the
analysis to more general matter couplings and to go beyond the semi-classical
approximation in the quantum theory. Since SIG coupled to matter has been shown
to be anomalous\cite{cangemi}, it is of interest to determine whether such
anomalies appear in the generic theory as well. Finally it is important
to understand the apparent relationship
between  black hole entropy and the imaginary part of the WKB phase.
\bigskip
\par\noindent
{\large\bf Acknowledgements}
\par
The authors are grateful to
A. Barvinsky, V. Frolov, J. Gegenberg, R.B. Mann and P. Sutton, for helpful
discussions.   This work
was supported in part by the Natural Sciences and Engineering
Research
Council of Canada.  \par
\end{document}